\documentclass[a4paper,12pt]{article}
\usepackage{amssymb}

\begin{document}

\title{Time-Variation of the Gravitational Constant and the Machian Solution
 in the Brans-Dicke Theory}
\author{A. Miyazaki \thanks{
Email: miyazaki@loyno.edu, miyazaki@nagasakipu.ac.jp} \vspace{3mm} \\
\textit{Department of Physics, Loyola University, New Orleans, LA 70118} \\
and \\
\textit{Faculty of Economics, Nagasaki Prefectural University} \\
\textit{Sasebo, Nagasaki 858-8580, Japan}}
\date{\vfill}
\maketitle

\begin{abstract}
The Machian cosmological solution satisfying $\phi =O(\rho /\omega )$ for
the perfect-fluid with negative pressure is discussed. When the coefficient
of the equation of state $\gamma \rightarrow -1/3$, the gravitational
constant approaches to constant. If we assume the present mass density $\rho
_{0}\sim \rho _{c}$ (critical density), the parameter $\epsilon $ ($\gamma
=(\epsilon -1)/3$) has a value of order $10^{-3}$ to support the present
gravitational constant. The closed model is valid for $\omega <-3/2\epsilon $
and exhibits the slowly accelerating expansion. We understand why the
coupling parameter $\left| \omega \right| $ is so large ($\omega \sim -10^{3}
$). The time-variation of the gravitational constant $\left| \dot{G}%
/G\right| \sim 10^{-13}\,yr^{-1}$ at present is derived in this model. 
\end{abstract}

\newpage

It has recently been cleared that the field equations of the Brans-Dicke
theory \cite{1)}\ does not necessarily produce those of general relativity
with the same energy-momentum tensor in the infinite limit of the coupling
parameter $\omega $ \cite{2)}, \cite{3)}. We have considered the physical
essence of the difference between general relativity and the Brans-Dicke
theory and have proposed the two postulates for the local and the
cosmological problems respectively \cite{4)}. The scalar field by
locally-distributed matter should exhibit the asymptotic behavior $\phi
=\left\langle \phi \right\rangle +O(1/\omega )$ for the large enough
coupling parameter $\omega $, and the scalar field of a proper cosmological
solution should have the asymptotic form $\phi =O(\rho /\omega )$ (the
Machian solution).

We have systematically surveyed the general existence of such Machian
cosmological solutions in the Brans-Dicke theory and have proved uniqueness
of the Machian solution for the homogeneous and isotropic universe with the
perfect-fluid matter with negligible pressure \cite{5)}. However, it is
unavoidable that the scalar field $\phi $\ goes to zero as the universe
expands satisfying the conservation law $a^{3}\rho =const$ in this
cosmological model. The time-variation of the gravitational constant in this
model is also not compatible with the recent observations (for examples \cite
{6)}, $\left| \dot{G}/G\right| \lesssim 1.6\times 10^{-12}\,yr^{-1}$).

We will discuss some alternatives of matter to explain this experimental
fact in the Machian point of view. First, we investigate simply the case of
the vacuum energy. The mass density $\rho _{m}$ of the perfect-fluid with no
pressure decreases gradually as the universe expands, and finally quantum
corrections to the vacuum must not become negligible in matter. This vacuum
energy density $\rho _{v}$ must almost be constant even though the universe
expands, and might keep the gravitational ''constant'' constant. Let us find
the Machian solution with $\phi =O(\rho /\omega )$ for the vacuum energy in
the Brans-Dicke theory.

The metric tensor for the homogeneous and isotropic universe is given as 
\begin{equation}
ds^{2}=-dt^{2}+a^{2}(t)[d\chi ^{2}+\sigma ^{2}(\chi )(d\theta ^{2}+\sin
^{2}\theta d\varphi ^{2})]\,,  \label{e1}
\end{equation}
where 
\begin{equation}
\sigma (\chi )\equiv \left\{ 
\begin{array}{l}
\sin \chi \;\;\;\;\;for\;k=+1\;(closed\;space) \\ 
\chi \;\;\;\;\;\;\;\;for\;k=0\;\ \;(flat\;space) \\ 
\sinh \chi \;\;\;for\;k=-1\;\ (open\;space)\,.
\end{array}
\right.  \label{e2}
\end{equation}
The source terms of the gravitational filed and the scalar field are the
energy-momentum tensor of the perfect-fluid with negligible pressure ($p=0$)
and the vacuum energy. The nonvanishing component of the energy-momentum
tensor is $T_{00}=-\rho c^{2}$ and the contracted energy-momentum tensor is $%
T=\rho c^{2}$, where the total density $\rho =\rho _{m}+\rho _{v}$. Let us
suppose that the mass density $\rho _{m}$ obeys independently the
conservation law $a^{3}\rho _{m}=const$ and the vacuum energy density $\rho
_{v}$ keeps constant. We discuss the epoch in which the relation $\rho
_{m}\ll \rho _{v}$ is satisfied after the universe expands enough: 
\begin{equation}
\rho =\rho _{v}=const\,.  \label{e3}
\end{equation}
The nonvanishing components of the field equations which we need solve
simultaneously are 
\begin{equation}
\frac{3}{a^{2}}\left( \dot{a}^{2}+k\right) =\frac{16\pi (1+\omega )}{%
(3+2\omega )c^{2}}\frac{\rho }{\phi }+\frac{\omega }{2}\left( \frac{\dot{\phi%
}}{\phi }\right) ^{2}+\frac{\ddot{\phi}}{\phi }  \label{e4}
\end{equation}
and 
\begin{equation}
\ddot{\phi}+3\frac{\dot{a}}{a}\dot{\phi}=\frac{8\pi }{(3+2\omega )c^{2}}\rho
\,,  \label{e5}
\end{equation}
where a dot denotes the partial derivative with respect to $t$.\ 

We seek Machian solutions satisfying $\phi =O(\rho /\omega )$. Let us
suppose that the scalar field $\phi $ is described as 
\begin{equation}
\phi (t)=\frac{8\pi }{(3+2\omega )c^{2}}\Phi (t)\,,  \label{e6}
\end{equation}
where an unknown function $\Phi (t)$ depends on only $t$ and should not
include the coupling parameter $\omega $\ in order that the scalar field $%
\phi (t)$ becomes Machian \cite{5)}. Substituting Eq.(\ref{e6}), we get from
Eq.(\ref{e5}) 
\begin{equation}
\ddot{\Phi}+3\frac{\dot{a}}{a}\dot{\Phi}=\rho \,,  \label{e7}
\end{equation}
which means that the ratio $\dot{a}/a$ also includes only $t$\ as the vacuum
energy density $\rho _{v}$ does not depend on $\omega $.\ So the expansion
parameter $a(t)$\ need have a form as 
\begin{equation}
a(t)\equiv A(\omega )\alpha (t)\,,  \label{e8}
\end{equation}
where $\alpha (t)$ is a function of only $t$\ and a coefficient $A(\omega )$
includes only $\omega $. Thus we obtain from Eq.(\ref{e4}) after eliminating 
$\ddot{\phi}$ by Eq.(\ref{e5}) 
\begin{equation}
\frac{\omega }{2}\left[ \left( \frac{\dot{\Phi}}{\Phi }\right) ^{2}+\frac{%
4\rho }{\Phi }\right] -\frac{3k}{A^{2}(\omega )\alpha ^{2}}=3\left( \frac{%
\dot{\alpha}}{\alpha }\right) ^{2}+3\left( \frac{\dot{\alpha}}{\alpha }%
\right) \left( \frac{\dot{\Phi}}{\Phi }\right) -\frac{3\rho }{\Phi }\,.
\label{e9}
\end{equation}

For the closed and the open spaces ($k=\pm 1$), if we require that Eq.(\ref
{e9}) is always satisfied for all arbitrary values of $\omega $, we find the
two following constraints must be held identically, 
\begin{equation}
\frac{\omega }{2}\left[ \left( \frac{\dot{\Phi}}{\Phi }\right) ^{2}+\frac{%
4\rho }{\Phi }\right] -\frac{3k}{A^{2}(\omega )\alpha ^{2}}\equiv C(t)
\label{e10}
\end{equation}
and 
\begin{equation}
3\left( \frac{\dot{\alpha}}{\alpha }\right) ^{2}+3\left( \frac{\dot{\alpha}}{%
\alpha }\right) \left( \frac{\dot{\Phi}}{\Phi }\right) -\frac{3\rho }{\Phi }%
\equiv C(t)\,,  \label{e11}
\end{equation}
where $C(t)$ is an arbitrary function of only $t$. From the constraint Eq.(%
\ref{e10}) for all arbitrary values of $\omega $, we obtain that\ the
coefficient $A(\omega )$ must have the following form\ 
\begin{equation}
\frac{3}{A^{2}(\omega )}=\left| \frac{\omega }{2}+B\right| \,,  \label{e12}
\end{equation}
where $B$\ is a constant with no dependence of $\omega $.

Let us adopt a notation that $j=-1$ for $\omega /2+B<0$ and $j=+1$ for $%
\omega /2+B>0$, for simplicity. Taking this notation and Eq.(\ref{e12}) into
account, we get from Eq.(\ref{e9}) 
\begin{equation}
\frac{\omega }{2}\left[ \left( \frac{\dot{\Phi}}{\Phi }\right) ^{2}+\frac{%
4\rho }{\Phi }-k\,j\frac{1}{\alpha ^{2}}\right] =3\left( \frac{\dot{\alpha}}{%
\alpha }\right) ^{2}+3\left( \frac{\dot{\alpha}}{\alpha }\right) \left( 
\frac{\dot{\Phi}}{\Phi }\right) -\frac{3\rho }{\Phi }+k\,j\frac{B}{\alpha
^{2}}\,.  \label{e13}
\end{equation}
We need to hold the two following identities to satisfy this equation for
all arbitrary $\omega $: 
\begin{equation}
\left( \frac{\dot{\Phi}}{\Phi }\right) ^{2}+\frac{4\rho }{\Phi }\equiv k\,j%
\frac{1}{\alpha ^{2}}  \label{e14}
\end{equation}
and 
\begin{equation}
3\left( \frac{\dot{\alpha}}{\alpha }\right) ^{2}+3\left( \frac{\dot{\alpha}}{%
\alpha }\right) \left( \frac{\dot{\Phi}}{\Phi }\right) -\frac{3\rho }{\Phi }%
\equiv -k\,j\frac{B}{\alpha ^{2}}  \label{e15}
\end{equation}
for $k=\pm 1$ and $j=\pm 1$, respectively.

We have a prospect of existence of solutions which have the following form: 
\begin{equation}
\Phi (t)=\zeta \rho (t)t^{2}\,,  \label{e16}
\end{equation}
\begin{equation}
\alpha (t)=bt\,,  \label{e17}
\end{equation}
where coefficients $\zeta $\ and $b$\ are constants respectively. In fact,
for $k\,j=+1$, we observe 
\begin{equation}
\zeta =1/8\,,\;\;b=1/6\,,\;\;B=5/12  \label{e18}
\end{equation}
satisfies Eqs.(\ref{e7}), (\ref{e14}), and (\ref{e15}). We can determine $%
\zeta $\ from Eq.(\ref{e7}), $b$ from Eq.(\ref{e14}), and $B$ from Eq.(\ref
{e15}) successively. It should be noted that no Machian solutions exist for
any combinations of $k\,j=-1$ in this case. If $\omega >-5/6$ for $k=+1$ or $%
-5/6>\omega >-2$ ($\omega \neq -3/2$) for $k=-1$, the gravitational force
becomes attractive ($G>0$). Equation (\ref{e16}) gives the decreasing
gravitational constant $G(t)\propto t^{-2}$ even if the vacuum energy
supports the constant mass density ($\rho _{v}=const$) in this Machian
solution. We can regard the vacuum energy as the cosmological constant $%
\Lambda $. If we observe the decaying cosmological constant $\Lambda
(t)\propto t^{-2}$, this means the vacuum energy density $\rho
_{v}(t)\propto t^{-2}$ and gives the constant scalar field $\Phi (t)=const$.

For the flat space case ($k=0$), the two identities are derived from Eq.(\ref
{e13}): 
\begin{equation}
\left( \frac{\dot{\Phi}}{\Phi }\right) ^{2}+\frac{4\rho }{\Phi }\equiv 0\,,
\label{e19}
\end{equation}
and 
\begin{equation}
\left( \frac{\dot{\alpha}}{\alpha }\right) ^{2}+\left( \frac{\dot{\alpha}}{%
\alpha }\right) \left( \frac{\dot{\Phi}}{\Phi }\right) -\frac{\rho }{\Phi }%
\equiv 0\,.  \label{e20}
\end{equation}
We find $\Phi (t)=-\rho _{v}t^{2}$ from Eq.(\ref{e19}) after integration (,
taking the integral constant to zero) and $\Phi (t)\alpha ^{2}(t)=const$
from Eq.(\ref{e19}) and (\ref{e20}), which gives $\alpha (t)\propto t^{-1}$.
It is obvious that these functions $\Phi (t)$, $\alpha (t)$ do not satisfy
Eq.(\ref{e7}) for $\rho _{v}=const$. No Machian solutions with the vacuum
energy $\rho _{v}=const$ exist for the flat space.

Next we discuss Machian solutions for the perfect-fluid with pressure $p$
(see also \cite{7)}-\cite{9)}), of which the energy-momentum tensor is
described as 
\begin{equation}
T_{\mu \nu }=-pg_{\mu \nu }-(\rho +p/c^{2})u_{\mu }u_{\nu }\,,  \label{e21}
\end{equation}
where $u^{\mu }$ is the four velocity $dx^{\mu }/d\tau $ ($\tau $ is the
proper time). The nonvanishing components are $T_{00}=-\rho c^{2}$, $%
T_{i\,i}=-pg_{i\,i}$ ($i\neq 0$), and its trace is $T=\rho c^{2}-3p$ for the
homogeneous and isotropic universe. The energy conservation $T_{;\nu }^{\mu
\nu }=0$ gives the equation of continuity 
\begin{equation}
\dot{\rho}+3\frac{\dot{a}}{a}\left( \rho +p/c^{2}\right) =0\,.  \label{e22}
\end{equation}
We suppose the equation of state 
\begin{equation}
p(t)=\gamma \rho (t)c^{2}\,,  \label{e23}
\end{equation}
where $0\leqq \gamma \leqq 1/3$ for the ordinary state. However, we will
neglect this constraint here and consider the wider range of $\gamma $ (at
least, $-1\leqq \gamma \leqq 1/3$). After integrating Eq.(\ref{e22}) with
the equation of state, we obtain 
\begin{equation}
\rho (t)\alpha ^{n}(t)=const\,,  \label{e24}
\end{equation}
where $n=3(\gamma +1)$. Equations (\ref{e4}) and (\ref{e5}) change to 
\begin{equation}
\frac{3}{a^{2}}\left( \dot{a}^{2}+k\right) =\frac{16\pi (1+\omega )}{%
(3+2\omega )c^{2}}\frac{\rho }{\phi }+\frac{\omega }{2}\left( \frac{\dot{\phi%
}}{\phi }\right) ^{2}+\frac{\ddot{\phi}}{\phi }+\frac{24\pi }{(3+2\omega
)c^{4}}\frac{p}{\phi }\,,  \label{eq24}
\end{equation}
\begin{equation}
\ddot{\phi}+3\frac{\dot{a}}{a}\dot{\phi}=\frac{8\pi }{(3+2\omega )c^{2}}%
\left( \rho -3p/c^{2}\right) \,,  \label{e25}
\end{equation}
respectively in this case. Taking Eqs.(\ref{e6}) and (\ref{e8}) into
account, we get directly after similar arguments 
\begin{equation}
\ddot{\Phi}+3\frac{\dot{\alpha}}{\alpha }\dot{\Phi}=\xi \rho  \label{e26}
\end{equation}
with $\gamma =(1-\xi )/3$ or $n=4-\xi $. After the elimination of $\ddot{\phi%
}$ from Eq.(\ref{eq24}) using Eq.(\ref{e25}), we find that Eq.(\ref{e13})
holds for the perfect-fluid with pressure as well as with no pressure.

For the flat space case ($k=0$), we find the same identities Eqs.(\ref{e19})
and (\ref{e20}) from Eq.(\ref{e13}) for all $\omega $, and then obtain 
\begin{equation}
\Phi (t)\alpha ^{2}(t)=const\,.  \label{e27}
\end{equation}
We have a prospect of existence of the following type of solution: 
\begin{equation}
\Phi (t)=\zeta \rho (t)t^{2}\,,  \label{e28}
\end{equation}
\begin{equation}
\alpha (t)=bt^{\beta }\,,  \label{e29}
\end{equation}
where $\beta $\ is a constant. We observe from Eqs.(\ref{e24}),(\ref{e26}),
and (\ref{e27}) 
\begin{equation}
\zeta =1/(\xi -5)\,,\;\;\beta =2/(2-\xi )\,,\;\;b:indefinite\,.  \label{e30}
\end{equation}
It is characteristic for the flat space that the coefficient $b$\ of the
expansion parameter $a(t)$ becomes indefinite. No other Machian solutions
exist in the range $0\leqq \xi <2$ for the flat space, because this solution
is continuous in this region\ for the continuous parameter $\xi $\ including 
$\xi =1$, for which the statement is proved \cite{5)}\ The coefficient $%
\zeta $ is negative for all $\xi $ ($0\leqq \xi \leqq 4$) or $n$ ($4\geqq
n\geqq 0$), so the gravitational force becomes attractive for $\omega <-2$.
The time-dependence of $\Phi (t)$ is described explicitly as 
\begin{equation}
\Phi (t)\propto t^{-4/(\xi -2)}\,.  \label{e31}
\end{equation}
The sign of the power reverses at $\xi =2$ or $n=2$. There are no parameters
to give a solution satisfying $\Phi (t)=const$.

For the closed and the open spaces ($k=\pm 1$), the equations which we need
solve simultaneously are Eqs.(\ref{e14}), (\ref{e15}), (\ref{e24}), and (\ref
{e26}). Similarly, we have a prospect of existence of the following type of
solution: 
\begin{equation}
\Phi (t)=\zeta \rho (t)t^{2}\,,  \label{e32}
\end{equation}
\begin{equation}
\alpha (t)=bt\,.  \label{e33}
\end{equation}
After calculating the power of $\Phi (t)$ in $t$ by Eqs.(\ref{e32}), (\ref
{e33}), and (\ref{e24}), we obtain 
\begin{equation}
\zeta =1/(\xi -2)  \label{e34}
\end{equation}
from Eq.(\ref{e26}), 
\begin{equation}
b=\left\{ 
\begin{array}{l}
(4-\xi ^{2})^{-1/2}\,,\;\;for\;k\,j=-1\;and\;0\leqq \xi <2 \\ 
(\xi ^{2}-4)^{-1/2}\,,\;\;for\;k\,j=+1\;and\;2<\xi \leqq 4\,
\end{array}
\right.  \label{e35}
\end{equation}
from Eq.(\ref{e14}), and 
\begin{equation}
B=-3/(\xi ^{2}-4)  \label{e36}
\end{equation}
from Eq.(\ref{e15}) successively. Thus, the Machian solution also exists in
these cases and unique for $0\leqq \xi <2$ because of continuity of the
parameter $\xi $\ from $\xi =1$ \cite{5)}. Though the parameter $\xi $\ is
constant for this solution, the same solution holds if $\xi $\ varies slowly
enough as the quasi-static process.

At the boundary ($\omega /2+B=0$) between the closed and the open spaces,
there exists a flat solution, which satisfies Eqs.(\ref{e14}), (\ref{e15}), (%
\ref{e24}), and (\ref{e26}) only for a particular value of the coupling
constant $\omega =6/(\xi ^{2}-4)$. In this solution, the coefficient $b$
becomes indefinite in the same way as the other cases for the flat space.
The parameter $\xi =1$ gives $\omega =-2$, which means $G=0$. At $\xi =0$
(for the universe with radiation), this solution becomes singular ($\omega
=-3/2$) and so we should discard it. We make the meaning of ''uniqueness'' 
\cite{5)} more definite by considering general cases for the perfect-fluid
with pressure.

The sign of the coefficient $\zeta $\ and the parameter $k\,j$\ reverse at $%
\xi =2$ or $n=2$ for the closed and the open spaces. To realize the
attractive gravitational force ($G>0$), we restrict to $\omega <-2$ for $%
k=+1 $ and $0\leqq \xi \leqq 1$, to $\omega <6/(\xi ^{2}-4)$ for $k=+1$ and $%
1<\xi <2$, and to $6/(\xi ^{2}-4)<\omega <-2$ for $k=-1$ and $1<\xi <2$.
Moreover, when $2<\xi \leqq 4$, we require $-2<\omega <6/(\xi ^{2}-4)$ for $%
k=+1$ and $6/(\xi ^{2}-4)<\omega $ for $k=-1$. In any cases, we exclude $%
\omega =-3/2$. The solution holds the type of $\Phi (t)\propto \rho (t)t^{2}$
and $a(t)\propto t$, and so the Machian relation 
\begin{equation}
\frac{G(t)M}{c^{2}a(t)}=const  \label{e37}
\end{equation}
is satisfied for all the time regardless of the time-dependence of the mass
density $\rho $.

In this Machian solution, the scalar field $\Phi (t)$ keeps almost constant
near $n=2$. Let us suppose that the present universe is described as the
case $n=2+\epsilon $ ($\epsilon \ll 1$), and then we get for the
time-dependence of $\Phi (t)$%
\begin{equation}
\Phi (t)=-(1/\epsilon )\rho (t)t^{2}\propto t^{-\epsilon }\,.  \label{e38}
\end{equation}
If we adopt $t_{0}/c\sim 10^{10}\,yr$ as the age of our universe and assume $%
\epsilon \sim 10^{-2}$, we find 
\begin{equation}
\left| \dot{\Phi}(t)/\Phi (t)\right| \propto \epsilon /(t/c)\sim
10^{-12}\,yr^{-1}\,,  \label{e39}
\end{equation}
which is compatible with the observational date for the time-variation of
the gravitational constant \cite{6)}.

The recent measurements \cite{10)} for the coupling parameter $\omega $\
gives a severe restriction $\left| \omega \right| \gtrsim 10^{3}$. Taking $%
\omega \sim -10^{3}$, $\epsilon \sim 10^{-2}$, $t_{0}/c\sim 10^{10}\,yr$,
and the present gravitational constant $G_{0}=6.67\times
10^{-8}\,dyn.cm^{2}.g^{-1}$ into account, we can estimate the present mass
density $\rho _{0}\sim 10^{-28}\,g.cm^{-3}$ from Eqs.(\ref{e6}) and (\ref
{e38}), which is ten times as large as the critical density $\rho _{c}\sim
10^{-29}\,g.cm^{-3}$. If we presume $\rho _{0}\sim \rho _{c}$, we obtain $%
\epsilon \sim 10^{-3}$ for the same other parameters, and its value gives $%
\left| \dot{G}/G\right| \sim 10^{-13}\,yr^{-1}$ at present.

The scalar field $\phi $\ has the asymptotic form $\phi =O(\rho /\omega )$
for the large coupling constant $\omega $\ in any cases discussed here, and
so the term $\omega (\dot{\phi}/\phi )^{2}$ appeared in the field equations
does not vanish in the infinite limit of $\omega $. The solution $\zeta
=-1/2 $, $b=1/2$, and $B=3/4$ for the closed or open universe with radiation
($\gamma =1/3$, $T=0$) does not exhibit the asymptotic behavior $\phi
=\left\langle \phi \right\rangle +O(1/\sqrt{\omega })$, though $T=0$, which
may be a counterexample to Faraoni \cite{3)}. The solution $\zeta =-1/5$, $%
\beta =1$ for the flat universe with radiation is also another
counterexample.

No Machian solutions for the flat space and for the closed or the open
spaces with $k\,j=-1$ in the case of the vacuum energy $\rho _{v}=const$.
The sign of $k\,j$ for the solution with the vacuum energy $\rho _{v}=const$
is opposite to that of the solution with the mass density $\rho =const$ ($%
\xi =4$, $n=0$). It should be noted that there is a discontinuity at $\xi =2$
and the sign of $k\,j$ ($\omega /2+B>0$ or $<0$) reverses there. The flat
solution for the perfect-fluid with pressure does not satisfy $\Phi
(t)=const $. The closed solution in the range of $0\leqq \xi <2$ seems to be
advantageous in the Machian point of view, taking the continuity of the
parameter $\xi $\ from $\xi =1$ and the sign of the gravitational constant.
The solution with $\xi =3$ or $\xi =4$ is not continuously connected with
that of $\xi =1$ as the quasi-static process of $\xi $.

The parameter $\xi =2$ which gives the Brans-Dicke scalar field $\phi
(t)=const$ means $\gamma =-1/3$, that is a ''negative'' pressure. This may
be mysterious, but recent measurements \cite{11)} of the distances to type
Ia supernovae have revealed that the expansion of the universe is rather
accelerating, which implies the existence of dark matter with negative
pressure. A slow varying scalar field with negative pressure ($-1<\gamma <0$%
) is recently known as \emph{quintessence} \cite{12)}. A cosmological
constant behaves like matter with negative pressure $\gamma =-1$ \cite{13)}.
So it does not seem to be necessarily unsound that we consider the existence
of matter with negative pressure $\gamma =-1/3$.

The expansion parameter for the closed space ($0\leqq \xi <2$) is explicitly
described as $a(t)=\left\{ -6/[\left( 4-\xi ^{2}\right) \omega +6]\right\}
^{1/2}\,t$ with the coupling parameter $\omega <-6/(4-\xi ^{2})$. For the
case $n=2+\epsilon $ ($\epsilon \sim 10^{-3}$), we get $a(t)\sim \lbrack
-3/(2\epsilon \omega +3)]^{1/2}\,t$ with $\omega <-3/2\epsilon $ for the
first order in $\epsilon $. Thus, if we require $\omega =-3/\epsilon $, we
find $a(t)\sim t$, and so we understand naturally the reason why the
coupling parameter $\left| \omega \right| $ is so large ($\omega \sim
-10^{3} $) at present. The expansion parameter $a(t)$ is a linear function
of $t$, but the parameter $\xi $\ approaches to $2$ very slowly as the
quasi-static process. Therefore, the expansion parameter increases a little
faster than $t $ and then the universe exhibits the accelerating expansion.

Equations.(\ref{e14}), (\ref{e15}), (\ref{e24}), and (\ref{e26}) are
invariant under the transformations $t\rightarrow t+t_{c}$, $t\rightarrow -t$%
, and $t\rightarrow t_{c}-t$ ($t_{c}$ is a positive constant) for the
solution Eqs.(\ref{e32}) and (\ref{e33}). So it is possible that this
solution describes the expansion or the collapse from a finite radius with a
finite gravitational constant.\newline
\newline
\textbf{Acknowledgment}

The author is grateful to Professor Carl Brans for helpful discussions and
his hospitality at Loyola University (New Orleans) where this work was done.
He would also like to thank the Nagasaki Prefectural Government for
financial support.

\end{document}